\documentclass[a4paper,fleqn,10pt]{article}
\usepackage[a4paper,text={13cm,20cm},centering,scale=0.7]{geometry}
\usepackage{amsfonts,amsmath,amsthm,amssymb,bm,graphicx,soul,ulem,xcolor}
\date{}
\begin{document}

\title{On the Jacobi metric for a general Lagrangian system\footnote{This article may be downloaded for personal use only. Any other use requires prior permission of the author and AIP Publishing. This article appeared in Journal of Mathematical Physics 60, 112901 (2019) and may be found at https://doi.org/10.1063/1.5124142.}}

\author{Paolo Maraner\footnote{Faculty of Economics and Management, Free University of Bozen-Bolzano,
Universit\"atsplatz - Piazza Universit\`a 1, 39100~Bozen-Bolzano, Italy. Email: \texttt{pmaraner@unibz.it}}}

\maketitle

%\date{October 31, 2019}

\begin{abstract} 
An explicit expression for the Jacobi metric for a general Lagrangian system is obtained as a series expansion in the square root of the kinetic energy of the system and the corresponding geodesics are described in terms of an appropriate non-linear connection and the associated curvature.
In the limit of low kinetic energies the trajectories of motion of any Lagrangian system are very well approximated by the geodesics of an energy dependent  Randers metric or, equivalently, by the paths in configuration space of a representative point subject to electromagnetic- and gravitational-like force fields. 
For higher kinetic energy values  the trajectories of motion are instead the geodesics of a general energy dependent Finsler metric, corresponding also to the paths in configuration space of a representative point subject to a hierarchy of a potentially infinite number of covariant force fields that generalise the electromagnetic and gravitational ones.
Some  general implications of these findings are discussed. \\

\end{abstract} 

\maketitle

\section{Introduction}
It has long been known \cite{Jacobi1842-43,Goldstein 1970,AKN 2006} that the \textit{paths} in configuration space of the representative point of a `natural' Lagrangian system, i.e.\! a mechanical system described by the Lagrangian 
\begin{equation}
\label{nL}
L(x,\dot{x})=\frac{1}{2}g_{ij}\dot{x}^i\dot{x}^j-V,
\end{equation}
where $x^i$, $i=1,...,d$  are coordinates on the configuration space $M$, dot indicates the differentiation with respect to the time $t$, 
$g_{ij}(x)$ is a Riemannian metric on $M$ 
and $V(x)$ is the potential energy of an external force field, are the geodesics of the energy dependent Jacobi metric
\begin{equation}
\label{ndsJ}
ds_{\!J}=\sqrt{2\left(E-V\right)g_{ij}dx^idx^j}\equiv\sqrt{j_{ij}dx^idx^j}.
\end{equation}
The Jacobi metric is obtained by eliminating time from the abbreviated (or Maupertuis) form of the action by means of the energy first integral and opens the way to the use of the methods of Riemannian geometry to the investigation of individual and collective properties of the paths in configuration space of a mechanical system \cite{Pin 1975,Krylov 1979,SHS 1996,Pettini 2010}. The parameterisation in terms of time is then recovered by observing that the arc length $s_{\!J}$ is related to the time $t$  by 
\begin{equation}
\label{dt}
dt=\frac{1}{2(E-V)}ds_{\!J}.
\end{equation}

The Jacobi metric generalises to the spatial paths of relativistic massive particles moving in static spacetimes \cite{Levi-Civita 1917,Weyl 1917,Gibbons 2016}. It further extends to the spatial paths of dynamical systems subject to velocity-dependent interactions, such as classical charged particles in a magnetic background or relativistic particles in a stationary spacetime \cite{Perlick 1991,CGGMW 2019}, with the important difference that the corresponding metric is no longer Riemannian, but a Finsler metric of Randers type \cite{Randers 1941}. More generally, in \cite{Mestdag2016} it has been proved that 
the solutions of the Euler-Lagrange equations of a strongly convex autonomous Lagrangian system are the geodesics of an associated energy dependent Finsler function \cite{Rund 1959,BCZ 2012}. 

In this paper we obtain an explicit expression for the Jacobi line element  for a general  Lagrangian that is analytic on the fiber of the configuration space tangent bundle, as well as differentiable on the base, as a series expansion in the square root of the kinetic energy of the system \cite{specific}.
We shall show that in the limit of low kinetic energies, the paths in configuration space of \textit{any} regular Lagrangian system are very well approximated by the geodesics of an energy dependent Randers metric. Correspondingly, in this regime, the paths of the system are very well approximated by those of a representative point moving on the configuration space under the simultaneous action of electromagnetic- and  gravitational-like force fields. For higher values of the kinetic energy the dynamics of a general Lagrangian system is instead associated with that of a representative point moving on the configuration space under the action of a hierarchy of a potentially infinite number of covariant force fields that are described by symmetric tensor potentials of arbitrary rank and that generalise the electromagnetic and gravitational ones. After comparing our findings with the previously known results, we investigate the geodesic equations associated with the general Jacobi line element. We note that the standard Finsler geometry formalism is not the most convenient for studying this class of Finsler spaces and we introduce an appropriate parameterisation of the geodesic equations as well as the corresponding non-linear connection and  curvature tensor.  Some general implications of our findings are discussed in the conclusive section.

\section{Lagrangian mechanics as Finsler geometry}

We consider a general Lagrangian system, a pair $(M, L)$ where the configuration space $M$ is a $d$-dimensional smooth manifold  parameterised by the coordinates $x^i$, $i=1,...,d$, and the Lagrangian $L$ is a smooth function on its tangent bundle $TM$. In the analysis of motion the Lagrangian is evaluated along the lift on the tangent bundle of a regular path $x(t)=(x^1(t),...,x^d(t))$ in $M$, so that $L$ is presented as a function of $x$ and $\dot{x}$, with $\dot{x}(t)=\frac{dx}{dt}(t)$.
 In the Jacobi formulation of Maupertuis' principle of least action, the path in configuration space of the representative point of the system  is obtained as a stationary point of the abbreviated action functional  
\begin{equation}
\label{aA}
 \int_{x_1}^{x_2} p_kdx^k,
\end{equation}
with the canonical momenta defined as usual by
$p_k(x,dx,dt)=\frac{\partial L}{\partial \dot{x}^k}(x,\dot{x})$ 
and the proviso that the quantity 
\begin{equation}
\label{first integral}
p_k\dot{x}^k-L\equiv E
\end{equation}
must be conserved on the varied path with fixed endpoints $x_1$ and $x_2$. In most applications the evolution parameter $t$ is identified with time so that the conserved quantity 
$E$ corresponds to the energy of the system, but this is not always necessarily the case. 
For  Lagrangian systems of type (\ref{nL}), Jacobi showed that upon elimination of 
$dt$
in (\ref{aA}) by means of  (\ref{first integral}), the paths of the system extremize the line integral of (\ref{ndsJ}), thus providing a completely geometrical characterisation of the system.

\subsection{Line element}
In order to extend this result to general Lagrangian systems, we will assume that  the Lagrangian $L(x,\dot{x})$ is sufficiently regular to be series expanded at $\dot{x}=0$, so that 
\begin{equation}
\label{L}
L(x,\dot{x})=l+l_i \dot{x}^i+\frac{1}{2}l_{ij}\dot{x}^i\dot{x}^j+...+\frac{1}{n!}l_{i_1...i_n}\dot{x}^{i_1}...\ \dot{x}^{i_n}+...
\end{equation}
with 
\begin{equation}
\label{ }
l_{i_1...i_n}(x)=\frac{\partial^n L}{\partial \dot{x}_{i_1}...\partial \dot{x}_{i_n}}(x,0)
\end{equation}
transforming as symmetric tensors on the configuration space $M$ for all $n\geq0$.
The canonical momenta of the system are thus obtained as
\begin{equation}
\label{pex }
p_k=\frac{\partial L}{\partial\dot{x}^k}=
l_k+ l_{ki}\dot{x}^i+...+
\frac{1}{(n-1)!}l_{ki_1...i_{n-1}}\dot{x}^{i_1}...\ \dot{x}^{i_{n-1}}+...,
\end{equation} 
and the abbreviated form of the action reads
\begin{equation}
\label{aAex}
\int_{x_1}^{x_2} p_k dx^k=\int_{x_1}^{x_2}\left(
l_i\dot{x}^i+ l_{ij}\dot{x}^i\dot{x}^j+...+
\frac{1}{(n-1)!}l_{i_1...i_{n}}\dot{x}^{i_1}...\ \dot{x}^{i_n}+...
\right)dt.
\end{equation}
The conserved quantity (\ref{first integral}) is correspondingly evaluated as 
\begin{equation}
\label{Eex}
E=-l+\frac{1}{2}l_{ij}\dot{x}^i\dot{x}^j
+...
+\frac{n-1}{n!}l_{i_1...i_n}\dot{x}^{i_1}...\ \dot{x}^{i_n}
+...
\end{equation}
Following the original derivation of Jacobi \cite{Jacobi1842-43, Goldstein 1970}, we can use this equation to obtain a formal expression of the differential $dt$ in terms of the coordinates $x^i$, the differentials $dx^i$ and the energy $E$, so that the evolution parameter $t$ can be fully eliminated from the abbreviated form of the action (\ref{aAex}), which will thus be expressed in purely geometrical terms.
In fact, the expression (\ref{Eex}) for the energy can be formally read as a series expansion of $E$ in the reciprocal of the differential $dt$. Upon series inversion we obtain
\begin{equation}
\label{dtauL}
\begin{split}
dt=&\ \sqrt{\frac{l_{ij}dx^i dx^j}{2(E+l)}}
+\frac{1}{3}\frac{l_{ijk}dx^i dx^j dx^k}{l_{ij}dx^i dx^j}
\\
&+\sqrt{2(E+l)}\left[\frac{1}{8}\frac{l_{ijkl}dx^i dx^j dx^k dx^l}{\left(l_{ij}dx^i dx^j\right)^\frac{3}{2}}-\frac{1}{6}\frac{\left(l_{ijk}dx^i dx^j dx^k\right)^2}{\left(l_{ij}dx^i dx^j\right)^\frac{5}{2}}\right]+...
\end{split}
\end{equation}
The substitution of (\ref{dtauL}) in (\ref{aAex}) yields  the abbreviated action $\int p_k(x,dx,dt)dx^k$  in the geometric form $\int ds_{\!J}(x,dx,E)$ with
\begin{equation}
\label{dsJ}
\begin{split}
ds_{\!J}=&\ l_i dx^i+\sqrt{2(E+l)l_{ij}dx^i dx^j}
+2(E+l)\frac{1}{6}\frac{l_{ijk}dx^i dx^j dx^k}{l_{ij}dx^i dx^j}
\\
&+[2(E+l)]^{\frac{3}{2}}\left[\frac{1}{24}\frac{l_{ijkl}dx^i dx^j dx^k dx^l}{\left(l_{ij}dx^i dx^j\right)^\frac{3}{2}}-\frac{1}{18}\frac{\left(l_{ijk}dx^i dx^j dx^k\right)^2}{\left(l_{ij}dx^i dx^j\right)^\frac{5}{2}}\right]+...
\end{split}
\end{equation}\\
For a fixed value of $E$ the Jacobi line element $ds_{\!J}(x,dx,E)$ is defined in the region of possible motion $M_E=\left\{x\in M : l\geq-E\right\}$, where is a homogeneous function of degree one of the differentials $dx^i$ 
\begin{equation}
\label{ }
ds_{\!J}(x,\lambda dx,E)=\lambda ds_{\!J}(x,dx,E)\hskip5pt\text{for all}\hskip5pt\lambda>0,
\end{equation} 
in accordance with what has been proven in \cite{Mestdag2016} (see also Appendix A).
If the necessary regularity  conditions are fulfilled, $ds_{\!J}(x,dx,E)$ describes a Finsler (or pseudo-Finsler) metric on $M_E$ \cite{Rund 1959,BCZ 2012}, fully characterised by the configuration space symmetric tensors $l_{i_1i_2...i_n}(x)$, with $n\geq0$. 

Since $ds_{\!J}$ appears as series expansion in $\sqrt{2(E+l)}$, for `small values' of $E+l$ the geometry associated with any regular Lagrangian system is very well approximated by a Randers type geometry. Correspondingly, in this regime the paths in configuration space of any regular Lagrangian system are very well approximated by those of a point moving on the manifold $M_E$, 
subject to gravitational- and electromagnetic-like interactions respectively described by the symmetric tensor potential $j_{ij}(x)=2[E+l(x)]l_{ij}(x)$ and  by the vector potential $l_i(x)$. 
The latter interaction dominates on the former.

It should be noted that this formal derivation of the Jacobi line element  is not a rigorous prove that the paths in configuration space of the general Lagrangian system (\ref{L}) are the geodesics of the energy dependent Finsler metric (\ref{dsJ}). This statement can be proved more simply \textit{a posteriori}, by showing that, with an appropriate choice of parameterisation, the geodesic equations associated with (\ref{dsJ}) are identical to the equations of motion for (\ref{L}). This will be shown in subsection \ref{Geodesic equations}.
An alternative derivation of the Jacobi line element (\ref{dsJ}) is given in Appendix B.

\subsection{Relation to known cases}
The Jacobi metric has so far been considered for `natural' Lagrangian systems \cite{Jacobi1842-43,Goldstein 1970,AKN 2006}, `natural' Lagrangian systems subject to magnetic-like interactions \cite{CGGMW 2019}, relativistic particles in a static spacetime \cite{Levi-Civita 1917,Weyl 1917,Gibbons 2016} and relativistic particles in stationary spacetimes \cite{Perlick 1991,CGGMW 2019}. In all these cases (\ref{dsJ}) reproduces the already known result.\\

For a \textit{`natural' Lagrangian system} described by the Lagrangian (\ref{nL}) the only non vanishing terms in (\ref{L}) are 
\begin{equation}
\label{ }
l=-V\hskip10pt\text{and}\hskip10pt l_{ij}=g_{ij},
\end{equation}
the evolution parameter $t$ is the Newtonian time, $E$ is the energy of the system and (\ref{dsJ})  coincides with the classical Jacobi metric (\ref{ndsJ}).\\

For a \textit{`natural' Lagrangian system subject to magnetic-like interactions} described by the vector potential $A_i(x)$, the Lagrangian reads
 \begin{equation}
\label{mnL}
L(x,\dot{x})=\frac{1}{2}g_{ij}\dot{x}^i\dot{x}^j+A_i\dot{x}^i-V
\end{equation}
where $g_{ij}(x)$ and $V(x)$ are again a metric and the potential energy of a force field on the configuration space $M$.
The only non vanishing terms in (\ref{L}) are then
\begin{equation}
\label{ }
l=-V,\hskip10pt l_i=A_i\hskip10pt\text{and}\hskip10pt l_{ij}=g_{ij},  
\end{equation}
and the line element (\ref{dsJ})  reads 
\begin{equation}
\label{}
ds_{\!J}=A_idx^i+\sqrt{2\left(E-V\right)g_{ij}dx^idx^j}.
\end{equation}
This is a Finsler  line element of Randers type.\\ 

For a \textit{relativistic particle in a static spacetime} with signature $+,-,-,-$, the Lagrangian reads
\begin{equation}
\label{staticst}
L(x,\dot{x})=-mc\sqrt{c^2g_{00}+g_{ij}\dot{x}^i\dot{x}^j},
\end{equation}
where $g_{00}(x)$ and $g_{ij}(x)$, $i,j=1,2,3$ are the non vanishing time independent components of the spacetime metric, the evolution parameter corresponds to the time coordinate and $E$ represents the relativistic energy of the particle. In the series expansion of (\ref{staticst}) all odd-order terms vanish identically, while the first few even-order terms are expressed in terms of $g_{00}(x)$ and $g_{ij}(x)$ as
\begin{equation}
\begin{split}
l=&-mc^2\sqrt{g_{00}},\hskip10pt
l_i=0,\hskip10pt
l_{ij}=-\frac{m}{\sqrt{g_{00}}}g_{ij},
\\
&l_{ijk}=0,\hskip10pt
l_{ijkl}=\frac{m}{c^2g_{00}^{3/2}}\left(g_{ij}g_{kl}+g_{ik}g_{jl}+g_{il}g_{jk}\right)
,\hskip10pt ...
\end{split}
\end{equation}
The line element (\ref{dsJ}) reduces then to 
\begin{equation}
ds_{\!J}=\left(\sqrt{2(E-mc^2\sqrt{g_{00}})}+\frac{[2(E-mc^2\sqrt{g_{00}})]^{\frac{3}{2}}}{8mc^2\sqrt{g_{00}}}+...\right)
\sqrt{-\frac{m}{\sqrt{g_{00}}}g_{ij}dx^idx^j},
\end{equation}
corresponding to the expansion in $\sqrt{2(E-mc^2\sqrt{g_{00}})}$ of the exact result  \cite{Gibbons 2016}.\\

For a \textit{relativistic particle in a stationary spacetime} with signature $+,-,-,-$ the Lagrangian reads instead 
\begin{equation}
\mathcal{L}=-mc\sqrt{c^2g_{00}+2cg_{0i}\dot{x}^i+g_{ij}\dot{x}^i\dot{x}^j},
\end{equation} 
with  $g_{00}(x)$,  $g_{0i}(x)$ and $g_{ij}(x)$, $i,j=1,2,3$, the time independent components of the spacetime metric and $t$ and $E$ again corresponding to the time coordinate and the relativistic energy respectively. The first few terms of the Lagrangian expansion (\ref{L}) read then
\begin{equation}
\begin{array}{l}\displaystyle
l=-mc^2\sqrt{g_{00}},\hskip10pt l_i=-\frac{mc}{\sqrt{g_{00}}}g_{0i},\hskip10pt 
l_{ij}=\frac{m}{\sqrt{g_{00}}}\gamma_{ij},
\\[15pt]\displaystyle\hskip10pt
l_{ijk}=-\frac{m}{c\sqrt{g_{00}^3}}\left(\gamma_{ij}g_{0k}+\gamma_{ki}g_{0j}+\gamma_{jk} g_{0i}\right),
\\[15pt]\displaystyle\hskip20pt
l_{ijkl}=\frac{m}{c^2 g_{00}^{3/2}}\Big(\gamma_{ij}\gamma_{kl}+\gamma_{ik}\gamma_{jl}+\gamma_{il}\gamma_{jk}+
\\[15pt]\displaystyle\hskip65pt
+2\frac{\gamma_{ij} g_{0k} g_{0l}}{ g_{00}}
+2\frac{\gamma_{ik} g_{0j} g_{0l}}{ g_{00}}
+2\frac{\gamma_{il} g_{0j} g_{0k}}{ g_{00}}
\\[15pt]\displaystyle\hskip70pt
+2\frac{\gamma_{jk} g_{0i} g_{0l}}{ g_{00}}
+2\frac{\gamma_{jl} g_{0i} g_{0k}}{ g_{00}}
+2\frac{\gamma_{kl} g_{0i} g_{0j}}{ g_{00}}
\Big),\hskip10pt  ...
\end{array}
\end{equation}
with $\gamma_{ij}=- g_{ij}+\frac{ g_{0i} g_{0j}}{ g_{00}}$ the spatial metric. The line element (\ref{dsJ}) reduces then to 
\begin{equation}
\label{dsJss}
\begin{split}
ds_{\!J}=& -\left(\frac{mc}{\sqrt{ g_{00}}} 
+\frac{2(E-mc^2\sqrt{g_{00}})}{2c g_{00}}\right)g_{0i}dx^i\\
&+\left(\sqrt{2(E-mc^2\sqrt{g_{00}})}+\frac{[2(E-mc^2\sqrt{g_{00}})]^{\frac{3}{2}}}{8mc^2\sqrt{g_{00}}}+...\right)
\sqrt{\frac{m}{\sqrt{g_{00}}}\gamma_{ij}dx^idx^j},
\end{split}
\end{equation}
again corresponding to the expansion in $\sqrt{2(E-mc^2\sqrt{g_{00}})}$ of the exact result \cite{CGGMW 2019}.

\subsection{Conformal rescaling}\label{cf}

As in the case originally considered by Jacobi, the scalar component of the Lagrangian's expansion, $l(x)=L(x,0)$, plays the role of the opposite of the potential energy of an external force filed acting on the system, while the remaining tensorial terms shape the geometry of the configuration space $M$.  In fact, on the one hand $l(x)$ is the only term that imposes restrictions on the definition of $ds_{\!J}(x,dx,E)$ and thus on the regions of possible motion. On the other hand, it can be fully reabsorbed in the tensors $l_{i_1i_2...i_n}(x)$ with $n>0$ by a conformal rescaling. By setting
\begin{equation}
\label{j}
j_{i_1i_2...i_n}=\left[2(E+l)\right]^{n-1}l_{i_1i_2...i_n}
\end{equation}
for $n\geq1$, the line element (\ref{dsJ}) takes  the  form 
\begin{equation}
\label{dsJj}
\begin{split}
ds_{\!J}=&\ j_i dx^i+\sqrt{j_{ij}dx^i dx^j}
+\frac{1}{6}\frac{j_{ijk}dx^i dx^j dx^k}{j_{ij}dx^i dx^j}
\\
&+\frac{1}{24}\frac{j_{ijkl}dx^i dx^j dx^k dx^l}{\left(j_{ij}dx^i dx^j\right)^\frac{3}{2}}-\frac{1}{18}\frac{\left(j_{ijk}dx^i dx^j dx^k\right)^2}{\left(j_{ij}dx^i dx^j\right)^\frac{5}{2}}+...
\end{split}
\end{equation}
showing that for each value of $E$, the geometry is fully characterised by the configuration space symmetric tensors $j_{i_1...i_n}(x,E)$, with $n\geq1$. The sum $E+l\equiv K$ can thus be identified with the kinetic energy of the system.

Under the conformal rescaling (\ref{j}), the differential of the evolution parameter (\ref{dtauL}) rewrites as 
\begin{equation}
\label{dtauL-dtauJ}
dt=\frac{1}{2(E+l)} d\tau_{\!J}
\end{equation}
with
\begin{equation}
\label{dtauJ}
\begin{split}
d\tau_{\!J}=&\ \sqrt{j_{ij}dx^i dx^j}
+\frac{1}{3}\frac{j_{ijk}dx^i dx^j dx^k}{j_{ij}dx^i dx^j}\\
&+\frac{1}{8}\frac{j_{ijkl}dx^i dx^j dx^k dx^l}{\left(j_{ij}dx^i dx^j\right)^\frac{3}{2}}-\frac{1}{6}\frac{\left(j_{ijk}dx^i dx^j dx^k\right)^2}{\left(j_{ij}dx^i dx^j\right)^\frac{5}{2}}+...
\end{split}
\end{equation}
Equation (\ref{dtauL-dtauJ}) generalises (\ref{dt}) to general Lagrangian systems,
with the difference that the differential  $dt$ is no longer a conformal rescaling of the Jacobi line element $ds_{\!J}$, but of the differential $d\tau_{\!J}=\frac{d}{d\ln \sqrt{2K}}ds_{\!J}$.

\subsection{Geodesic equations}\label{Geodesic equations}
The paths in configuration space of a  general Lagrangian system can be obtained by extremizing the line integral of  (\ref{dsJj})  
and therefore correspond to the geodesics of a Finsler (or pseudo-Finsler) metric. 
The methods and results of Finsler geometry \cite{Rund 1959,BCZ 2012} can thus be  applied to the investigation of a general Lagrangian system. Nonetheless, the explicit form of $ds_{\!J}(x,dx,E)$  in terms of the configuration space tensors $j_{i_1...i_n}(x,E)$ suggests that this particular type of Finsler geometry can be handled directly by tensor analysis on $M_E$, like Riemannian geometry, rather than by the complicated tangent bundle formalism of Finsler geometry \cite{Chern 1996}.
With this in mind, we proceed by exploiting homogeneity to introduce an arbitrary parameterisation of the line integral (\ref{aA}), 
\begin{equation}
\label{li}
\int_{x_1}^{x_2} ds_{\!J}(x,dx,E)=
\int_{x_1}^{x_2} ds_{\!J}\left(x,\acute{x} ,E\right)d\vartheta,
\end{equation}
where the acute accent indicates differentiation with respect to an arbitrary parameter $\vartheta$. 
The choice of  parameter is important, because the form of the geodesic equations depends on it. In Riemannian geometry the most convenient choice is the metric arc length, for which the geodesic equations take their familiar form. In Finsler geometry is customary to make the same choice, so that the geodesic equations can be rearranged in a form identical to those of Riemannian geometry, with a `metric tensor' and the corresponding `Christoffel symbols' explicitly depending on the $ \acute{x}$. However, this simplicity is only apparent, as is possible to see by explicitly writing down the geodesic equations even for the simplest Finsler geometry. On the other hand, the direct variation of (\ref{li}) produces the geodesic equations in the form  
\begin{equation}\label{gge}
\begin{split}
&\Gamma_{hi}\acute{x}^i +
\\ &\hskip10pt 
+\frac{1}{\acute{\tau}_J}\left(j_{hi}\acute{\acute{x}}^i+\Gamma_{hij}\acute{x}^i\acute{x}^j\right)-\frac{\acute{\acute{\tau}}_J}{\acute{\tau}_J^2}j_{hi}\acute{x}^i+
\\ &\hskip30pt  
+\frac{1}{\acute{\tau}_J^2}\left(j_{hij}\acute{\acute{x}}^i\acute{x}^j+\Gamma_{hijk}\acute{x}^i\acute{x}^j\acute{x}^k\right)-\frac{\acute{\acute{\tau}}_J}{\acute{\tau}_J^3}j_{hij}\acute{x}^i\acute{x}^j+
\\ &\hskip40pt 
+\frac{1}{\acute{\tau}_J^3}\left(\frac{1}{2}j_{hijk}\acute{\acute{x}}^i\acute{x}^j\acute{x}^k+\Gamma_{hijkl}\acute{x}^i\acute{x}^j\acute{x}^k\acute{x}^l\right) -\frac{1}{2}\frac{\acute{\acute{\tau}}_J}{\acute{\tau}_J^4}j_{hijk}\acute{x}^i\acute{x}^j\acute{x}^k+...=0,
\end{split}
\end{equation}
with $\tau_{\!J}$ the Jacobi parameter defined in (\ref{dtauJ}) and the quantities $\Gamma_{hi_1i_2...i_n}(x,E)$ defined as
\begin{eqnarray}
\label{Gammas}
\begin{split}
\Gamma_{hi} =&\ \partial_i j_h-\partial_h j_i, 
\\
\Gamma_{hij}  =&\ \frac{1}{2}\left(\partial_i j_{hj}
+\partial_j j_{ih}-\partial_h j_{ij}\right), \\
\Gamma_{hijk}  =&\ \frac{1}{3!}\left(\partial_i j_{hjk}
+\partial_j j_{ihk}+\partial_k j_{ijh}
-\partial_h j_{ijk}\right),  
\\
\Gamma_{hijkl}  =&\ \frac{1}{4!}\left(\partial_i j_{hjkl}
+\partial_j j_{ihkl}+\partial_k j_{ijhl}+\partial_l j_{ijkh}
-\partial_h j_{ijkl}\right), \\
&  ...
\end{split}
\end{eqnarray}
The geodesic equations (\ref{gge}) take their simpler form when $\vartheta$ is chosen equal to the Jacobi parameter $\tau_{\!J}$ and not to the metric arc length $s_{\!J}$. In fact, by setting $d\vartheta\equiv d\tau_{\!J}$ we obtain the equations in the form
\begin{equation}
\label{ge}
\begin{split}
&\left(j_{hi}+j_{hij}{x^j}'+\frac{1}{2}j_{hijk}{x^j}'{x^k}'+...\right){x^i}''
\\ &\hskip20pt
+\Gamma_{hi}{x^i}'+\Gamma_{hij}{x^i}'{x^j}'+\Gamma_{hijk}{x^i}'{x^j}'{x^k}'+\Gamma_{hijkl}{x^i}'{x^j}{x^k}'{x^l}'+...=0,
\end{split}
\end{equation}
with the prime symbol indicating differentiation with respect to $\tau_{\!J}$. 
This choice of parameterisation explicitly breaks the invariance under the rescaling in the ${x}'$, but has other advantages.
On the one hand, equations (\ref{ge}) appear as a natural generalisation of the familiar covariant form of the Riemannian geodesic equations on $M_E$, to which they reduce when all $j_{i_1,...i_n}(x,E)$ except $j_{ij}(x,E)$ vanish identically. On the other hand, they are closely related to the equations of motion of the original Lagrangian system, to which they reduce under the conformal rescaling of the evolution parameter (\ref{dtauL-dtauJ}).  The quantities $\Gamma_{hi_1...i_n}(x,E)$ appear as a generalisation of the Christoffel symbols and are all constructed by the same rule from the symmetric tensors $j_{i_1...i_n}(x,E)$.
Up to an overall sign,  $\Gamma_{hi}(x,E)$ is the exterior derivative of the one-form $j_i(x,E)$ and transforms as a rank two antisymmetric tensor; $\Gamma_{hij}(x,E)$ are the Christoffel symbols of the first kind associated with the metric tensor $j_{ij}(x,E)$; the remaining $\Gamma_{hi_1...i_n}(x,E)$ with $n>2$, display non-covariant transformation rules somehow analogues to that of the Christoffel symbols and, to the best of our knowledge, have never been considered before in the literature. 

In the regions where $j_{ij}(x,E)$ is everywhere non-degenerate, equations (\ref{ge}) can be resolved with respect to the second derivatives. The existence and uniqueness for their solutions follow then from standard ODE-theory. Denoting as usual by $j^{ij}(x,E)$ the inverse of $j_{ij}(x,E)$, $j^{ik}j_{kj}=\delta^i_j$, we proceed by contracting both terms of (\ref{ge}) by the inverse of $j_{hi}+j_{hij}{x^j}'+\frac{1}{2}j_{hijk}{x^j}'{x^k}'+...$, the rank two contravariant tensor
\begin{equation}
\label{ }
j^{gh}-{j^{gh}}_i{x^i}'-\left(\frac{1}{2}{j^{gh}}_{ij}
-{j^g}_{ik}{j^{hk}}_j\right){x^i}'{x^j}'+...,
\end{equation}
where indices are raised by means of $j^{ij}(x,E)$. After renaming indices, the geodesic equations (\ref{ge}) take then the form
\begin{equation}
\label{contrage}
{x^h}''
+{\gamma^h}_i{x^i}'+{\gamma^h}_{ij}{x^i}'{x^j}'+{\gamma^h}_{ijk}{x^i}'{x^j}'{x^k}'+...=0
\end{equation}
with 
\begin{equation}
\label{gammas}
\begin{split}
{\gamma^h}_i=&\ {\Gamma^h}_i\\
{\gamma^h}_{ij}=&\ {\Gamma^h}_{ij}-{j^h}_{k(i}{\Gamma^k}_{j)}\\
{\gamma^h}_{ijk}=&\ {\Gamma^h}_{ijk}-{j^h}_{l(i}{\Gamma^l}_{jk)}
+{j^h}_{l(i}{j_j}^{lm}{\Gamma^n}_{k)}j_{mn}
-\frac{1}{2}{j^h}_{l(ij}{\Gamma^l}_{k)}\\
& ...
\end{split}
\end{equation}
and where round brackets indicate symmetrisation. 
Since the ${\gamma^h}_{ij}(x,E)$ transform as the components of a connection  on $M_E$,  the sum ${x^h}''+{\gamma^h}_{ij}{x^i}'{x^j}'$  transforms as a tensor  and so do all remaining terms of the expansion. In particular, all ${\gamma^h}_{i_1...i_n}(x,E)$ with $n\neq2$ transform as tensors on $M_E$. This may be made explicit by observing that the non-covariant quantities $\Gamma_{hi_1...i_n}(x,E)$ always appear in the (\ref{gammas}) through the covariant combinations 
\begin{equation}
\label{F}
\begin{split}
F_{hi_1...i_n}&={\Gamma}_{hi_1...i_n}-\frac{1}{(n-2)!}j_{hj(i_1...i_{n-2}}{\Gamma^j}_{i_{n-1}i_n)}\\
&=\frac{1}{n!}\left(\nabla_{i_1}j_{hi_2...i_{n}}+...+\nabla_{i_1}j_{i_1...i_{n-1}h}
-\nabla_{h}j_{i_1...i_{n}}\right),
\end{split}
\end{equation}
where $\nabla_i$ indicates the covariant derivative associated with the Levi-Civita connection ${\Gamma^h}_{ij}$.
As a whole, equations (\ref{contrage}) appear then as the equation of motion of a particle subject to an electromagnetic-like interaction described by the vector potential $j_i(x,E)$ and the force filed $F_{ij}(x,E)\equiv\Gamma_{ij}(x,E)$, a gravitational-like interaction described by the tensor potential $j_{ij}(x,E)$ and the force filed $\Gamma_{hij}(x,E)$ and to a hierarchy of a potentially infinite number of a  new type of interactions described by the symmetric tensor potentials $j_{i_1...i_n}(x,E)$ and the force fields $F_{hi_1...i_n}(x,E)$ and straightforwardly generalising the electromagnetic and gravitational interactions.

\subsection{Non-linear connection and curvature}
For each value of $E$, equations (\ref{contrage}) can be seen as the auto parallel equations 
\begin{equation}
\label{}
{x^h}''+{\boldsymbol{\gamma}^h}_i {x^i}'=0
\end{equation}
of a non-linear connection ${\boldsymbol{\gamma}^h}_i(x,{x}',E)$ on the base manifold $M_E$ \cite{Rund 1959,BCZ 2012,Minguzzi 2014}, with
\begin{equation}
\label{ }
{\boldsymbol{\gamma}^h}_i={\gamma^h}_i+{\gamma^h}_{ij}{x^j}'+{\gamma^h}_{ijk}{x^j}'{x^k}'+...
\end{equation}
The connection is determined only up to an arbitrary torsion term ${T^h}_{ij}(x,{x}',E){x^j}'$, with  ${T^h}_{ij}=-{T^h}_{ji}$, and is one of the many inequivalent non-linear connections that can be associated with the Finslerian line element (\ref{dsJj}). It is worth noticing that  ${\boldsymbol{\gamma}^h}_i(x,x',E)$ is not the standard Cartan non-linear connection generally considered in Finsler geometry, but rather  a generalisation of the non-linear connection considered in \cite{Miron 2004} for Randers spaces. 

The curvature tensor ${\boldsymbol{r}^h}_{ij}(x,x',E)$ associated with the non-linear connection ${\boldsymbol{\gamma}^h}_i(x,x',E)$ is as usual given by 
\begin{equation}
\label{ }
{\boldsymbol{r}^h}_{ij}=
\partial_i {\boldsymbol{\gamma}^h}_j-
\partial_j {\boldsymbol{\gamma}^h}_i-
{\boldsymbol{\gamma}^k}_i\frac{\partial{\boldsymbol{\gamma}^h}_j}{\partial {x^k}'}+
{\boldsymbol{\gamma}^k}_j\frac{\partial{\boldsymbol{\gamma}^h}_i}{\partial {x^k}'}.
\end{equation}
This is still expressed as a series in $\dot{x}$ as 
\begin{equation}
\label{ }
{\boldsymbol{r}^h}_{ij}={r^h}_{ij}+{r^h}_{ijk}{x^k}'+{r^h}_{ijkl}{x^k}' {x^l}'+...
\end{equation}
with the first ${r^h}_{i_1...i_n}(x,E)$ given by
\begin{equation}
\label{curvature components}
\begin{split}
{r^h}_{ij}=&\ \nabla_i {F^h}_j-\nabla_j {F^h}_i
-\frac{1}{2}{j^h}_{ik}{F^k}_l{F^l}_j+\frac{1}{2}{j^h}_{jk}{F^k}_l{F^l}_i\\
{r^h}_{ijk}=&\ {R^h}_{ijk}-
\nabla_j{j^h}_{l(k}{F^l}_{i)}+
\nabla_k{j^h}_{l(j}{F^l}_{i)}
-2{F^l}_j{\gamma^h}_{kli}+2{F^l}_k{\gamma^h}_{jli}\\
&\ -{j^l}_{m(j}{F^m}_{i)}{j^h}_{n(k}{F^n}_{l)}
+{j^l}_{m(k}{F^m}_{i)}{j^h}_{n(j}{F^n}_{l)}\\
& ...
\end{split}
\end{equation}
where ${R^h}_{ijk}=\partial_j{\Gamma^h}_{ki}-\partial_k{\Gamma^h}_{ji}- {\Gamma^l}_{ji}{\Gamma^h}_{kl}+{\Gamma^l}_{ki}{\Gamma^h}_{jl}$ is the Riemann tensor associated with the Levi-Civita connection ${\Gamma^h}_{ij}$. As in Riemannian geometry it is possible to construct a contracted curvature tensor $\boldsymbol{r}_i(x,x',E)$, the analogue on the Ricci tenor $R_{ij}={R^h}_{ihj}$, as
\begin{equation}
\label{ }
\boldsymbol{r}_i={\boldsymbol{r}^h}_{ih}=
r_i+r_{ij}{x^j}'+r_{ijk}{x^j}'{x^k}'+...
\end{equation}
The first few $r_{i_1...i_n}(x,E)$ are obtained by contraction of the (\ref{curvature components}). 
It is also possible to construct a curvature scalar $\boldsymbol{r}(x,x',E)$ 
by further contraction with ${x^i}'$, 
\begin{equation}
\label{cs}
\boldsymbol{r}=\boldsymbol{r}_i{x^i}'.
\end{equation}
However, this should not be confused with the analogue of the scalar curvature $R=j^{ij}R_{ij}$ of Riemannian geometry, whose generalisation is less straightforward.

\section{Discussion and Conclusions}

A classical result of Jacobi, describing the paths in configuration space
of a `natural' Lagrangian system as the geodesics of an energy dependent Riemannian metric, has been extended by showing that the trajectories of motion of a general Lagrangian system are the geodesics of an energy dependent Finsler metric. 
An explicit expression of the Finslerian line element has been obtained in terms of a potentially infinite number of symmetric tensors on the configuration space and provides a natural generalisation of the Riemannian line element.
Quite remarkably, in the limit of low kinetic energies the paths in configuration space of \textit{any} regular Lagrangian system are very well approximated by the geodesics of a Randers metric. 
This general result naturally leads to speculations, on the one hand, on 
the necessity of type and structure of classical fundamental interactions and, on the other hand, on the effectiveness of Riemannian geometry.\\

The type and structure of classical fundamental interactions are in fact encoded in the Lagrangian that describes the motion of a test particle.  For a test particle, the configuration space is the physical spacetime parameterised by the time coordinate $x^0$ and the spatial coordinates $x^1$, $x^2$ and $x^3$. The evolution parameter $t$ is arbitrary and does not coincide with either the time coordinate $x^0$ or the physical time. Correspondingly, the conserved quantity $E$ is no longer the energy of the system, but some other constant characterising the particle. 
We have shown in full generality that the paths in configuration space of a system of this type are the geodesics of the line element (\ref{dsJ}) with the scalar $l(x)$ equal to zero if no external forces act, as is natural to assume in this case. In the limit of small values of $E$ the action for the test particle is therefore proportional to the line integral of the first few terms of (\ref{dsJ}) with $l(x)\equiv0$ and is thus given by
\begin{equation}
\label{ }
 S\approx-\alpha\int\Big( l_i dx^i+\sqrt{2E}\sqrt{l_{ij}dx^idx^j}\Big)
\end{equation}
with $l_i(x)$ and $l_{ij}(x)$ symmetric tensors on spacetime and $\alpha$  a constant characterising the particle. In the context of Lagrangian mechanics this structure is  universal.
The only information that needs to be added is the dimension of spacetime, the non-degeneracy and signature of $l_{ij}(x)$ and the values of $E$ and $\alpha$.
A comparison with the relativistic action for a charged particle and the relative strength of electromagnetic and gravitational interactions,  provides the values
\begin{equation}
\label{ }
\sqrt{2E}=\frac{mc}{\alpha}=\frac{m}{e}\sqrt{4\pi\epsilon_0G},
\end{equation}
where $m$ is the mass of the particle, $e$ its charge, $\epsilon_0$ the electric constant and $G$ the gravitational constant. For an electron $\sqrt{2E}\approx10^{-21}$, largely justifying the assumption of small values of $E$.
The dynamics of the electromagnetic potential $A_i(x)=\frac{c^2}{\sqrt{4\pi\epsilon_0G}}l_i(x)$ and of the gravitational potential $g_{ij}(x)=l_{ij}(x)$
is not determined a priori, but is curious to observe that by setting the curvature scalar
(\ref{cs}) equal to zero,  we obtain exactly the free Maxwell and Einstein equations. It is tempting to speculate that these considerations could have anticipated electromagnetic and gravitational force fields. \\ 
The general structure of the line element (\ref{dsJ}) also suggests two different, but not exclusive, ways of extending the classical theory of fields. 
On the one hand, we can investigate the effect of a universal `external' scalar potential, letting $l(x)$ be different from zero. Our experience of spacetime is local, where this field could be approximately constant, after all.
On the other hand, we can proceed to a purely Finslerian extension by taking into account the higher rank symmetric tensors  $l_{i_1...i_n}(x)$ for $n>2$. Given the smallness of the expansion parameter their effect is certainly extremely weak, but it might become relevant when integrated over very large spacetime distances. Even if fully geometrical in nature, this extension can be presented as a theory of the covariant fields $F_{hi_1...i_n}(x)$ for $n\neq2$ in the Riemannian background provided by the metric $l_{ij}(x)$. In this respect the theory  is radically different from the Finslerian unification attempts of the past \cite{Goenner 2004,Goenner 2014} and from the more recently proposed Finslerian spacetime theories \cite{Beem 1970,Asanov 1985,Rutz 1993,Pfeifer and Wohlfarth 2011,Pfeifer and Wohlfarth 2012,Laemmerzahl and Perlick 2018}. 
   
The second speculation concerns the effectiveness of (pseudo-)Rienannian geometry 
in the description of space(time). At present the building block of geometry is the Pythagorean distance formula: it is assumed on infinitesimal scales, extended to arbitrary signatures, transformed to arbitrary coordinates and eventually postulated as the fundamental line element in (pseudo-)Riemannian geometry. In the light of the considerations above, this chain of implications is reversed with the Riemannian line element that does not need to be postulated. In fact, if we accept that our perception of geometry comes from physical experience and thus, in last analysis, from the observation of the motion of test particles, in the low energy limit in which we live we are necessarily led to a Randers type line element. This necessarily reduces to a line element of Riemannian type  when the effect of the much stronger electromagnetic-like forces is cancelled by the formation of electrically neutral `atoms' and `molecules'. The (pseudo-)Riemannian line element emerges universally from regular Lagrangians with non-singular  Hessian matrix $l_{ij}(x)=\frac{\partial L}{\partial \dot{x}^i\dot{x}^j}(x,0)$. 
 If we add that the set of Lagrangians  with a singular  Hessian matrix   forms a subset of zero measure of the set of all regular Lagrangians, we are lead to conclude, with Riemann, that the investigation of the next simplest case 
``would be rather time-consuming and throw proportionally little new light on the study of space" \cite{gijkl}.  In our perspective, non-Pythagorean geometries represent a subset of zero measure of the set of all possible geometries.

In conclusion, we must mention that the geometry associated with a general Lagrangian system has been extensively  studied in the literature as a generalisation of Finsler geometry \cite{Kern 1974,Miron and Anastasiei 1994,Bucataru and Miron 2007}. 
The general argument presented in \cite{Mestdag2016} and the present study show instead that the geometry of smooth manifolds endowed with a regular Lagrangian can be studied entirely within Finsler geometry.

\appendix
\section*{Appendix A: The Jacobi metric as a Finsler metric}\label{AppA}
In \cite{Mestdag2016}  Routh reduction has been used to show that the paths in configuration space of a strongly convex autonomous Lagrangian system are the geodesics of an associated energy dependent Finsler metric. Here, we present an elementary, though purely formal, argument to show that the Jacobi line element associated with a general Lagrangian system is always homogeneous of degree one in the differentials $dx^i$, corresponding thus to a general Finsler metric in the broadest sense of the term.
To see this, we recall that in order to calculate the Jacobi metric associated with a general autonomous Lagrangian system one can use the  conservation of energy 
\begin{equation}
\label{H}\textstyle
H\left(x,\frac{dx}{dt}\right)\equiv E
\end{equation}
to formally express the differential $dt$ in terms  of the coordinates $x^i$, the differentials $dx^i$ and the energy $E$  and substitute it in the abbreviated form of the action $\int p_idx^i$. Since the equation implicitly defining $dt$ as a function of the $x^i$, $dx^i$  and $E$ depends on the differentials $dx^i$ only through the quantities $\frac{dx^i}{dt}$, implicit differentiation implies that 
\begin{equation}
\label{ }
dx^i\frac{\partial dt(x,dx,E)}{\partial dx^i}=dt(x,dx,E).
\end{equation}
From the converse of Euler's homogeneous function theorem \cite{Lindeloef1932} it follows then that  $dt(x,dx,E)$ is a homogeneous function of degree one of the $dx^i$. Correspondingly $\frac{dx^i}{dt(x,dx,E)}$ is a homogeneous function of degree zero of the $dx^i$ and so are the momenta $p_i(x,\frac{dx}{dt})$ that depend on $dt$ only through the quantities $\frac{dx^i}{dt}$. Accordingly,  the Jacobi line element 
\begin{equation}
\textstyle
ds_{\!J}=p_i\left(x,\frac{dx}{dt(x,dx,E)}\right)dx^i
\end{equation}
is a homogeneous function of degree one of $dx^i$.  No more general geometric structure than Finsler geometry is required for the description of regular autonomous Lagrangian systems.
\section*{Appendix B: A second derivation of the Jacobi line element}\label{B}
The Jacobi line element  for a general Lagrangian system that  can be expanded as a power series in the $\dot{x}^i$,  can also be derived by the method used in \cite{Gibbons 2016} and \cite{CGGMW 2019} to obtain the Jacobi metric for timelike geodesics in static and stationary spacetimes respectively.
The idea is to use the conservation of energy to rewrite the Jacobi Lagrangian, i.e. the integrand of the abbreviated form of the action (\ref{aAex}),
\begin{equation}
\label{JL}
L_{J}=l_i\dot{x}^i+l_{ij}\dot{x}^i\dot{x}^j
+\frac{1}{2}l_{ijk}\dot{x}^i\dot{x}^j\dot{x}^k
+\frac{1}{6}l_{ijkl}\dot{x}^i\dot{x}^j\dot{x}^k\dot{x}^l
+...,
\end{equation}
as a homogeneous function of degree one in the $\dot{x}^i$. The Jacobi line element $ds_{\!J}$ can then be read directly from the identity $ds_{\!J}=\frac{ds_{\!J}}{dt}dt=L_Jdt$. To this end, we rewrite (\ref{JL}) as 
\begin{equation}
\label{JL2}
L_{J}=l_i\dot{x}^i+\sqrt{l_{ij}\dot{x}^i\dot{x}^j}\sqrt{ l_{ij}\dot{x}^i\dot{x}^j}
+\frac{1}{2}l_{ij}\dot{x}^i\dot{x}^j\frac{l_{ijk}\dot{x}^i\dot{x}^j\dot{x}^k}{l_{ij}\dot{x}^i\dot{x}^j}
+\frac{1}{6}\left(l_{ij}\dot{x}^i \dot{x}^j\right)^\frac{3}{2}\frac{l_{ijkl}\dot{x}^i\dot{x}^j\dot{x}^k\dot{x}^l}{\left(l_{ij}\dot{x}^i \dot{x}^j\right)^\frac{3}{2}}
+...
\end{equation}
and observe that the conservation of energy  (\ref{Eex}) yields
\begin{equation}
\label{lij}
l_{ij}\dot{x}^i\dot{x}^j=2(E+l)-\frac{2}{3}l_{ijk}\dot{x}^i\dot{x}^j\dot{x}^k-\frac{1}{4}l_{ijkl}\dot{x}^i\dot{x}^j\dot{x}^k\dot{x}^l+...
\end{equation}
The substitution of (\ref{lij}) in the first factor of each term of (\ref{JL2}), the expansion of the square root and the rearrangement of each term taking into account  again (\ref{JL2}), eventually yields the Jacobi Lagrangian in the form 
\begin{equation}
\label{ }
\begin{split}
L_J=&l_i\dot{x}^i+\sqrt{2(E+l)l_{ij}\dot{x}^i\dot{x}^j}
+2(E+l)\frac{1}{6}\frac{l_{ijk}\dot{x}^i\dot{x}^j\dot{x}^k}{l_{ij}\dot{x}^i\dot{x}^j}\\
&+\left[2(E+l)\right]^{\frac{3}{2}}\left[\frac{1}{24}\frac{l_{ijkl}\dot{x}^i\dot{x}^j\dot{x}^k\dot{x}^l}{\left(l_{ij}\dot{x}^i \dot{x}^j\right)^\frac{3}{2}}-\frac{1}{18}\frac{\left(l_{ijk}\dot{x}^i\dot{x}^j\dot{x}^k\right)^2}{\left(l_{ij}\dot{x}^i \dot{x}^j\right)^\frac{5}{2}}\right]+...,
\end{split}
\end{equation}
confirming (\ref{dsJ}) as the correct Jacobi line element.

%%%%%%%%%%%%%% 

%%%%%%%%%%%%%%


\begin{thebibliography}{99}
%
\bibitem{Jacobi1842-43}
C. G. J. Jacobi, \textit{Vorlesungen \"uber dynamik} (G. Reimer Verlag, Berlin, 1866). 
%
\bibitem{Goldstein 1970}
H. Goldstein, \textit{Classical Mechanics} (Addison-Wesley, Reading Massachusetts, 1970).
%
\bibitem{AKN 2006}
V. I. Arnold, V. V. Kozlov and A. I. Neishtadt, \textit{Mathematical aspects of classical and celestial mechanics} (Springer Verlag, Berlin, 2006);
https://doi.org/10.1007/978-3-540-48926-9.
%
\bibitem{Pin 1975}
O. C. Pin, 
Curvature and Mechanics,
Adv. Math. 15, 269 (1975);
https://doi.org/10.1016/0001-8708(75)90139-5.
%
\bibitem{Krylov 1979}
N. S. Krylov, \textit{Works on the Foundations of Statistical Physics} (Princeton University Press, New Jersey, 1979)
%
\bibitem{SHS 1996}
M. Szydlowskii, M. Heller and W. Sasin, 
Geometry of spaces with the Jacobi metric,
J. Math. Phys.  37, 346 (1996);
https://doi.org/10.1063/1.531394.
%
\bibitem{Pettini 2010}
M. Pettini, \textit{Geometry and topology in Hamiltonian Dynamics and Statistical Mechanics} (\textit{Interdisciplinary Applied Mathematics} vol 33) (Springer Verlag, Berlin, 2010);
https://doi.org/10.1007/978-0-387-49957-4.
%
\bibitem{Levi-Civita 1917}
T.~Levi-Civita,
Statica einsteiniana, 
Rend. R. Accad. Linc. 26, 458 (1917).
%
\bibitem{Weyl 1917}
H. Weyl, 
Zur Gravitationstheorie,
Ann. Phys. 359, 117 (1917);
https://doi.org/10.1002/andp.19173591804.
%
\bibitem{Gibbons 2016}
G. W. Gibbons,
The Jacobi metric for timelike geodesics in static spacetimes,
Class. Quantum Grav. 33, 025004 (2016);
https://doi.org/10.1088/0264-9381/33/2/025004.
%
\bibitem{Perlick 1991}
V. Perlick,
The brachistochrone problem in a stationary space-time,
J. Math, Phys. 32, 3148 (1991);
https://doi.org/10.1063/1.529472.
%
\bibitem{CGGMW 2019}
S. Chanda, G. W. Gibbons, P.  Guha, P. Maraner and M. W. Werner,
Jacobi-Maupertuis Randers-Finsler metric for curved spaces and the gravito-electro-magnetic effect, J. Math. Phys. 60, 122501 (2019);
https://doi.org/10.1063/1.5098869.
%
\bibitem{Randers 1941}
G. Randers, 
On an asymmetrical metric in the four-space of general relativity, 
Phys. Rev. 59, 195 (1941);
https://doi.org/10.1103/PhysRev.59.195.
%
\bibitem{Mestdag2016}
T. Mestdag, 
Finsler geodesics of Lagrangian systems through Routh reduction,
Mediterr. J. Math. 13, 825 (2016); 
https://doi.org/10.1007/s00009-014-0505-z
%
\bibitem{Rund 1959}
H. Rund, \textit{The Differential Geometry of Finsler Spaces} (Springer Verlag, Berlin, 1959);
https://doi.org/10.1007/978-3-642-51610-8.
%
\bibitem{BCZ 2012}
D. Bao, S.-S. Chern and Z. Shen, \textit{An introduction to Riemann-Finsler geometry}  (Springer Verlag, Berlin, 2012);
https://doi.org/10.1007/978-1-4612-1268-3.
%
\bibitem{specific}
Throughout this paper we do not discuss the convergence of this series. All considered series are implicitly assumed to be convergent. Moreover, the term `Finsler geometry' is used in its broadest sense, as a geometry described by a line element homogeneous of degree one in the differentials $dx^i$.
%
\bibitem{Chern 1996}
S.-S. Chern,
Finsler Geometry Is Just Riemannian Geometry without the Quadratic Restriction, 
Notices Am. Math. Soc. 43, 959 (1996).
%
\bibitem{Minguzzi 2014}
E. Minguzzi, 
The connections of pseudo-Finsler spaces, 
Int. J. Geom. Methods Mod. Phys. 11, 1460025 (2014);
https://doi.org/10.1142/S0219887814600251.
%
\bibitem{Miron 2004}
R. Miron, 
The geometry of Ingarden spaces, 
Rep. Math. Phys. 54,  131 (2004);
https://doi.org/10.1016/S0034-4877(04)80010-7.
%
\bibitem{Goenner 2004}
H. F. M. Goenner,
On the history of unified field theories, 
Living Rev. Relativ. 7, 2 (2004);
https://doi.org/10.12942/lrr-2004-2.
%
\bibitem{Goenner 2014}
H. F. M. Goenner,
On the history of unified field theories. Part II. (ca. 1930-ca. 1965), 
Living Rev. Relativ. 17, 5 (2014);
https://doi.org/10.12942/lrr-2014-5.
%
\bibitem{Beem 1970}
J. K. Beem, 
Indefinite Finsler spaces and timelike spaces, 
Can. J. Math. 22, 1035 (1970);
https://doi.org/10.4153/CJM-1970-119-7.
%
\bibitem{Asanov 1985}
G. S. Asanov, \textit{Finsler Geometry, Relativity and Gauge Theories} (Springer Netherlands, 1985),
https://doi.org/10.1007/978-94-009-5329-1.
%
\bibitem{Rutz 1993}
S. Rutz, 
A Finsler generalisation of Einstein's vacuum field equations, 
Gen. Relativ. Gravit. 25, 1139 (1993);
https://doi.org/10.1007/BF00763757.
%
\bibitem{Pfeifer and Wohlfarth 2011}
C. Pfeifer and M. N. R. Wohlfarth,
Casual structure of electrodynamics on Finsler spacetimes, 
Phys. Rev. D 84, 044039 (2011);
https://doi.org/10.1103/PhysRevD.84.044039.
%
\bibitem{Pfeifer and Wohlfarth 2012}
C. Pfeifer and M. N. R. Wohlfarth,
Finsler geometric extension of Einstein gravity, 
Phys. Rev. D 85, 064009 (2012);
https://doi.org/10.1103/PhysRevD.85.064009.
%
\bibitem{Laemmerzahl and Perlick 2018}
C. L\"ammerzahl and V. Perlick, 
Finsler geometry as a model for relativistic gravity, 
Int. J. Geom. Methods Mod. Phys. 15, 1850166 (2018);
https://doi.org/10.1142/S0219887818501669.
%
\bibitem{gijkl}
If we repeat the inversion of the series (\ref{Eex}) and the subsequent substitution of $dt$ in (\ref{aAex}) under the assumption that $l_{ij}(x)$ vanishes identically, we obtain that the first even order term in the expansion of the line element $ds_{\!J}$ is proportional the fourth root of a differential expression of the fourth degree, as foreseen by Riemann.
%
\bibitem{Kern 1974}
J. Kern,
Lagrange geometry, 
Arch. Math. 25, 438 (1974);
https://doi.org/10.1007/BF01238702.
%
\bibitem{Miron and Anastasiei 1994}
R. Miron and M. Anastasiei \textit{The Geometry of Lagrange Spaces: Theory and Applications} (Springer Netherlands, 1994);
https://doi.org/10.1007/978-94-011-0788-4.
%
\bibitem{Bucataru and Miron 2007}
I. Bucataru I and R. Miron R, \textit{Finsler-Lagrange geometry: Applications to dynamical systems} (Editura Academiei Romane, Buchurest, 2007).
%
\bibitem{Lindeloef1932}
E. Lindel\"of, \textit{Differentiali- ja integralilasku ja sen sovellutukset II} (Mercator Printing Company Limited, Helsinki,1932) 
\end{thebibliography}
\end{document}